\documentclass[a4paper,prb,twocolumn]{revtex4}
\usepackage{amsmath}
\usepackage{graphicx}
\usepackage{subfigure}

\usepackage{color}

\def \be{\begin{equation}\begin{aligned}}
\def \ee{\end{aligned}\end{equation}}
\def \c{c^{\dagger}}
\def \la{\langle}
\def \ra{\rangle}
\def \i{{\textrm i}}

\begin{document}
\title{Particle-hole symmetric localization in optical lattices \\ using time
modulated random on-site potentials}
\author{Yue Zou${}^1$, Ryan Barnett${}^2$, Gil Refael${}^1$}
\affiliation{${}^1$Department of Physics, California Institute of
Technology, Pasadena, California 91125, USA \\ ${}^2$Joint Quantum Institute and Condensed Matter Theory Center, \\
 Department of Physics, University of Maryland, College Park, MD 20742, USA}
\begin{abstract}
{The} random hopping models exhibit many
fascinating features, such as diverging localization length and
density of states as energy approaches the bandcenter, due to its
particle-hole symmetry. Nevertheless, such models {are yet to be}
realized {experimentally} because the particle-hole symmetry is easily
destroyed by diagonal disorder. Here we propose that a pure random
hopping model can be {effectively} realized in {ultracold atoms} by
modulating {a} disordered onsite potential
{in particular frequency ranges}. {This idea is motivated by
  the} recent development of the phenomena called ``dynamical
localization'' or ``coherent destruction of tunneling''. Investigating
the application of this idea in one dimension, we find that if the
oscillation frequency of the {disorder} potential is gradually
increased from zero to infinity, one can tune a non-interacting system
from an Anderson insulator to a random hopping model with diverging
localization length at the band center, and eventually to a
uniform-hopping tight-binding model.

\end{abstract}
\date{\today}

\maketitle

\section{Introduction}

The interplay of quenched disorder and low dimensionality in quantum
systems has been a central theme {in} condensed matter
physics. A beautiful theory describes how the spectrum of disordered
Hamiltonians follows from the symmetries of the problem
\cite{Zirnbauer1996}. 
In this work, we focus on {the} one-dimensional non-interacting random
hopping model, namely {the} tight-binding model with {pure}
off-diagonal disorder.  {This model's} properties have been extensively
investigated theoretically using various techniques for many
years. This model {also exhibits physics manifest in several} other models, such as
quantum particles connected by random strength springs, spin 1/2
random XX chain{s}, random quantum Ising chain{s} in {a} transverse field, and
random mass Dirac fermions. This model exhibits many surprising
features. Early theoretical
work\cite{Dyson,Thouless1972,Theodorou1976,Eggarter1978} focused on
properties derivable from the mean local Green's function, notably the
typical localization length which diverges as $\sim{\ln |E|}$ and the
mean density of states (DOS) also diverging as $\sim1/{|E(\ln
E^2)^3|}$ as energy $E$ approaches the band-center.  {Such behavior is}
very different from Anderson insulators where disorder appears in the
diagonal terms of the Hamiltonian, and there are no singularities in the localization
length or density of states spectrum. More recent work {has} studied
this model using real{-}space RG\cite{Fisher1994} and supersymmetry
methods\cite{Balents1997} and has uncovered more interesting results,
most importanly, an additional length scale - mean localization length
as a function of energy - 
which diverges as $\sim{\ln^2 |E|}$. Most recently, the effect of
random hopping amplitudes on interacting fermionic and bosonic systems has
also been investigated. In the 2d fermionic case\cite{Gade,Motrunich,Foster2008}, it has been
shown that random hopping amplitude, on top of nontrivial spectral effects, could lead to a novel type of
instability; in the bosonic case\cite{Gil2004,Gil2008}, a ``Mott glass''
phase has been predicted in addition to usual Mott insulating and
superfluid phases.

On the other hand, the experimental realization of a pure random
hopping model has proven to be extremely difficult. This is mainly because diagonal
disorder inevitably arises when we try to disorder bond values, and any amount of diagonal disorder
would break the particle-hole symmetry of the random-hopping model and
thereby destroy the universal behavior of this class of disordered
Hamiltonians.

In this work, we propose that a pure random hopping model can be
realized in optical lattices by first creating an Anderson insulator
and then {modulating} the disordered on-site potential energy
periodically. Our idea is closely connected to recent work on the
phenomena dubbed ``dynmaical localization'' or ``coherent destruction
of tunneling'' in double wells, semiconductor superlattices, and
recently optical
lattices\cite{Dunlap1986,Grossmann1991,Holthaus1992,Ignatov1993,Holthaus1995,Meier1995,Grifoni,Holthaus2001,Eckardt2005,Eckardt2005b,Creffield2006,Martinez2006r,Martinez2006,Eckardt2007,Lignier2007,Kierig2008,Luo2008,Tsukada2008,Kayanuma,Creffield2009,Eckardt2009}. While
one can use this path to the random hopping model in any
dimensionality ($d\le 3$), we will concentrate on its 1d application
below.

The basic idea of dynamical localization is the following. Consider a
particle in a double-well potential with a tunneling amplitude $J$ and
a time-{modulated} potential energy 
offset:
\be
H=-J(a^{\dagger}b+b^{\dagger}a)+V\cos(\omega t)(a^{\dagger}a-b^{\dagger}b).
\ee
By performing a unitary transformation with
\be 
U=e^{-i\frac{V}{\omega}\sin(\omega t)(a^{\dagger}a-b^{\dagger}b)},
\ee
one can readily obtain that in the large-$\omega$ limit, the original Schr\"odinger equation is transformed to the effective Hamiltonian
\be
H_{eff}=-J\mathcal{J}_0\left(\frac{V}{\omega}\right)(a^{\dagger}b+b^{\dagger}a),
\ee
where $\mathcal J_0$ is the zeroth Bessel function. Thus one can see
that the effect of time-modulating the potential energy is {to} simply
renormalize the tunneling amplitude $J$ in the large-$\omega$
limit. This phenomena has been observed in
experiments\cite{Lignier2007,Kierig2008,Eckardt2009}{.}  {It}  has 
{also been} proposed
{as} a method to tune interacting bosons through {the}
superfluid-insulator transition\cite{Eckardt2005,Creffield2006}, to
observe the analog of photon assisted tunneling and Shapiro
steps\cite{Ignatov1993,Eckardt2005b}, and to manipulate the
localization properties of Anderson
insulators\cite{Holthaus1995,Martinez2006r}. Recent experimental
work has confirmed some of these proposals \cite{Sias,Zenesini}.

\begin{figure}
\includegraphics[scale=0.4]{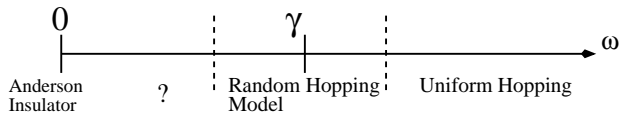}\caption{Phase diagram of
  the model (\ref{model}) studied in this work. At zero frequency, the
  system is an Anderson insulator; when the frequency $\omega$ is
  comparable to the disorder width $\gamma$, the system behaves as a
  random hopping model; when $\omega$ is much larger than $\gamma$,
  the system crosses over to the uniform-hopping tight-binding regime,
  which is fully achieved when $\omega=\infty$. We also present some
  interesting and puzzling results for the regime $0<\omega\ll\gamma$.}\label{phase_diagram}
\end{figure}

For our purpose, it suffices to notice that the original potential energy $V$ resides
in the renormalization factor of the hopping amplitude. Thus, if one
modulates an Anderson insulator instead, one expects that the disorder
in the onsite energy should be transformed into the disorder of
hopping amplitude in the same way. In other words, one obtains the
random hopping model by {rapidly} modulating the disordered
potential energies of an Anderson insulator. As we will see
{in} later sections, however, if the modulating frequency $\omega$ is much
larger than typical potential energy, this randomness in effective
hopping amplitude is suppressed, and we obtain a uniform-hopping
tight-binding model. Therefore, the random hopping model behavior
survives when the frequency $\omega$ is comparable to the typical
potential energy. In summary, as the oscillation frequency of the
potential energy is gradually increased from zero to infinity, one can
tune a non-interacting system from an Anderson insulator to a random
hopping model with diverging localization length at the band center,
and eventually to a uniform-hopping tight-binding model (see
Fig. \ref{phase_diagram}).

Note, also, that a different but related model was studied in
Refs. (\onlinecite{Holthaus1995,Martinez2006r,Martinez2006}), where an
Anderson insulator with stationary disordered potentials but
oscillating linear potential (e.g., uniform AC electric field) is
considered. In that model, the localization properties of the Anderson insulator
can be manipulated by the oscillating linear potential, but the random
hopping model behavior is not accessible.

In the {remainder} of this work, we will analyze the localization properties of the following model:
\be\label{model}
H &= H_0 + 2 V \cos(\omega t),\\
H_0 &= - J \sum_{n=1}^{N-1}(\c_nc_{n+1}+\c_{n+1}c_n),\\
V&=\sum_{n=1}^Nv_n\c_nc_n,
\ee
where $N$ is the system size, and we assume the onsite potential $v_n$
obeys a uniform distribution between $[-\gamma/2,\gamma/2]$. In
Sec. \ref{sec:Floquet}, we will introduce the Floquet formalism with
which we obtain the localization length and the density of states of
the time-dependent Hamiltonian (\ref{model}). In Sec. \ref{sec:Heff},
an effective time-independent Hamiltonian is shown to emerge from the
high-frequency limit of the original time-dependent Hamiltonian
(\ref{model}) in analogy to the dynamical localization
phenomena. Numerical results from both the Floquet calculation and the
effective Hamiltonian calculation are presented and discussed in
detail in Sec. \ref{sec:result}. Next, we discuss possible
experimental methods to modulate on-site disorder and to detect
signatures of random hopping models in optical lattices. Finally, we
summarize our results in Sec. \ref{sec:summary}.

\section{Computation of The Density of States And The Localization Length}\label{sec:Floquet}

For a time-periodic system with Hamiltonian $H(t)$ and period
$T=2\pi/\omega$, by the Floquet theorem, 
its wavefunctions {can be written in the form} 
\be\label{psi_phi}
\psi(t)=e^{-\i Et}\phi(t), 
\ee 
where $E$ 
is the quasienergy defined {modulo} 
$\omega$, and $\phi(t+T)=\phi(t)$. {Here and in what follows we
  have set $\hbar=1$.}
{This well-known result is the analog of the
Bloch's theorem for particles in a periodic spatial potential.}
To solve for $E$ and
$\phi(t)$, one approach is {to} rewrite the Schrodinger equation
\be
\i\partial_t\psi(t)=H\psi(t)
\ee
as
\be
H_F\phi=E\phi,
\ee
where $H_F$ is the so-called Floquet Hamiltonian
\be
H_F = H -\i \partial t
\ee
which is a matrix in the augmented Hilbert space $\mathcal H \times
\mathcal T$, where $\mathcal H$ is the original Hilbert space, and
$\mathcal T$ is the frequency space\cite{Shirley, Sambe1973}, and then
to find the {eigenvalues} and eigenstates of $H_F$. Alternatively, it is
also well-known that $e^{-iET}$ and $\phi(T)$ are the eigenvalue and
eigenstate of the Floquet operator
\be
\mathcal F=\mathcal T \exp\left(-\i\int_0^T\textrm{d}tH(t)\right),
\ee
where $\mathcal T$ is the time-ordering operator. 

To obtain the density of states, we work with the latter
approach. First, we calculate the Floquet operator $\mathcal F$ by
{the} numerical Trotterization procedure. Then, we diagonalize the
Floquet operator $\mathcal F$ to find the quasienergies which we
define to be in the ``first Brillouin zone'' $-\omega/2\leq
E\leq\omega/2$. Then, we obtain the cumulative distribution function
of the quasienergies, average it over many realization of disorder,
numerically differentiate it with respect to quasienergy, and finally
obtain the density of states.

We would also like to obtain the localization length of this model for
arbitrary frequency $\omega$. For one-dimensional non-interacting
time-independent {systems} with $N$-sites, we recall that the
localization length of a state with energy $E$ is given by
\cite{Thouless1972}
\be\label{thouless}
\frac1{\lambda(E)} = - \lim_{N\rightarrow\infty} \frac1N \ln |G_{1N}(E)|,
\ee
where the Green's function
\be
G(E) = (E I- H)^{-1},
\ee
$I$ is the identity matrix.

Following Ref. \onlinecite{Martinez2003,Martinez2006}, we generalize the concept of localization length of a time-periodic system by defininng it as the localization length of the time-averaged wavefunction. In terms of the Green's function, it is 
\be\label{Floquet_loc}
\frac1{\lambda(E)} = - \lim_{N\rightarrow\infty} \frac1N \ln|G_{1N}(E)|
\ee
where 
\be
G=\la \Omega=0 | G_F(E) | \Omega=0  \ra
\ee
and $G_F$ is the so-called Floquet Green's function:
\be
G_F(E) = (EI - H_F)^{-1}.
\ee
Here, if we denote the frequency operator $\hat{\Omega}=\i\partial_t$ and its eigenstates
\be
\hat{\Omega}|n\ra=n\omega|n\ra,
\ee
$|\Omega=0\ra$ introduced above is simply the eigenstate with $n=0$.

Next, we discuss how to compute $G\equiv\la \Omega=0 | G_F(E) | \Omega=0  \ra$. For a Hamiltonian of the form 
\be
H &= H_0 + 2 V \cos(\omega t),
\ee
from
\be
(EI-H_F)G_F=I,
\ee
we insert the {resolution} of identity in the frequency space and obtain
\be
\sum_p\la m|(EI-H_F)|p\ra\la p|G_F|n\ra=I\la m|n\ra.
\ee
{Thus},
\be
\left[(E+m\omega)I-H_0\right]G_{mn}-VG_{m+1,n}-VG_{m-1,n}=I\delta_{mn},\nonumber
\ee
where
\be
G_{mn}\equiv\la m |G| n \ra.
\ee
To solve $G_{00}(E)$ from this system of equations, we follow
Ref. \onlinecite{Martinez2006} to obtain
\be\label{Green}
G(E)\equiv G_{00}(E) = (EI - H_0 - V_{eff}^+ - V_{eff}^- )^{-1},
\ee
where
\be
V_{eff}^{\pm}=V\frac1{E\pm1\omega-H_0-V\frac1{E\pm2\omega-H_0-V\frac1{\vdots}V}V}V.
\ee
{The} number of iterations needed to ensure the convergence of $V_{eff}^{\pm}$ is roughly proportional to $\gamma/\omega$.

Finally, it would be interesting to mathematically check if the
Thouless relation\cite{Thouless1972} between the density of states and
the localization length holds for a Floquet system.

\section{Effective Hamiltonian in The Fast Oscillation Limit }\label{sec:Heff}

In this section, we show that if the oscillation frequency $\omega$ is comparable or larger than the disorder width $\gamma$, the original Schrodinger equation of a time-dependent Hamiltonian can be transformed to that of a time-independent effective Hamiltonian. For the original Schrodinger equation
\be
\i\partial_t\psi=H\psi,\qquad
H=H_0+2V\cos(\omega t),
\ee
we define 
\be
\psi = U\tilde{\psi},\qquad
U = e^{-2\i\sin(\omega t)V/\omega}.
\ee
Then the {Schrodinger} equation becomes
\be
\i\partial_t\tilde{\psi}=H_{eff}\tilde{\psi},\qquad
H_{eff} &= U^{\dagger}HU-2V\cos(\omega t).\nonumber
\ee
Using the operator identity
\be
e^{\i\eta\c_nc_n}c_ne^{-\i \eta\c_nc_n} = c_ne^{-\i\eta},
\ee
we have
\be
H_{eff}
&=-J\sum_n\left[(\c_nc_{n+1}+\c_{n+1}c_n)\right.\\
&\times\sum_{m=-\infty}^{\infty}(-1)^m\mathcal J_m\left(\frac{2(v_n-v_{n+1})}{\omega}\right)\cos(m\omega t)\\
&+\i(\c_nc_{n+1}-\c_{n+1}c_n)\\
&\times\left.\sum_{m=-\infty}^{\infty}(-1)^{m+1}\mathcal J_m\left(\frac{2(v_n-v_{n+1})}{\omega}\right)\sin(m\omega t)\right].\nonumber
\ee
For $\omega$ larger than or comparable to $\gamma$, the argument of
the Bessel functions is comparable {to} or smaller than 1. {Hence}
$\mathcal J_0$ dominates over other Bessel functions, and we obtain an
effective time-independent Hamitonian
\be\label{Heff}
H_{eff}^{(0)}\approx-J\sum_n\left(\c_nc_{n+1}+\c_{n+1}c_n\right)\mathcal J_0\left(\frac{2(v_n-v_{n+1})}{\omega}\right),
\ee
which is a tight-binding model with a site-dependent effective hopping amplitude
\be\label{Jeff}
J_{eff,n}=J\mathcal J_0\left(\frac{2(v_n-v_{n+1})}{\omega}\right).
\ee

When $\omega$ is not too large, this is exactly the random hopping model we are looking for, and it should exhibit behaviors such as diverging localization length and density of states at the band center. We can compute the localization length of this effective Hamiltonian by using (\ref{thouless}) and compare with the exact calculation using (\ref{thouless}):
\be\label{Geff}
\frac1{\lambda(E)} &= - \lim_{N\rightarrow\infty} \frac1N \ln |G_{1N}(E)|,\\
G_{eff}(E) &= (EI-H_{eff}^{(0)})^{-1}.
\ee
It is also straightforward to compute the density of states of this model numerically.

However, when $\omega\gg\gamma$,
\be
\mathcal J_0\left(\frac{2(v_n-v_{n+1})}{\omega}\right)\approx1
\ee 
regardless of the value of $v_n$. {Therefore} in this limit the
system behaves like a uniform-hopping tight-binding model (see
FIG. \ref{phase_diagram}). In this regime, we expect the localization
length of every state {to diverge}, and the density of states diverges at
the band edges instead.

\begin{figure}
\centering
\includegraphics[scale=0.65]{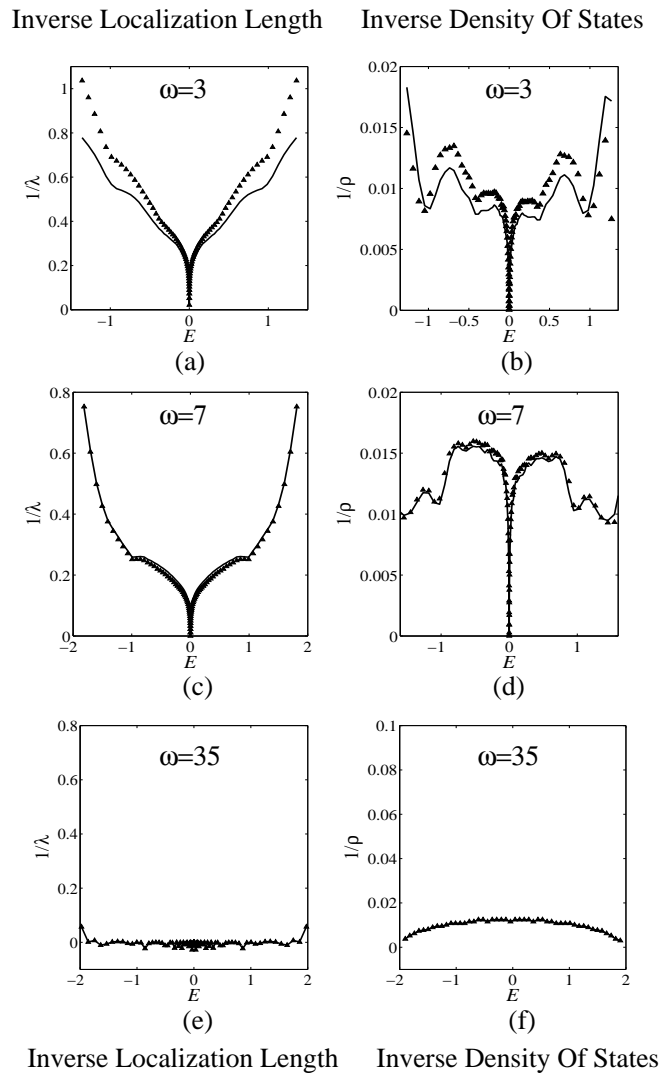}
\caption{Inverse of the localization length $1/\lambda$ (a, c, e) and inverse of the density of states $1/\rho$ (b, d, f) vs. quasienergy $E$ (in units of the hopping amplitude $J$). Disorder width $\gamma=10J$, hopping amlitude $J=1$, oscillation frequency $\omega=3J, 7J, 35J$. Solid lines are from the Floquet calculation of the original model (\ref{model}); solid triangles are from the effective Hamiltonian $H_{eff}^{(0)}$ (\ref{Heff}). Averaged over 1000 realizations of disorder.}\label{fig:compare}
\end{figure}

\section{Numerical results}\label{sec:result}

\begin{figure}
\includegraphics[scale=0.4]{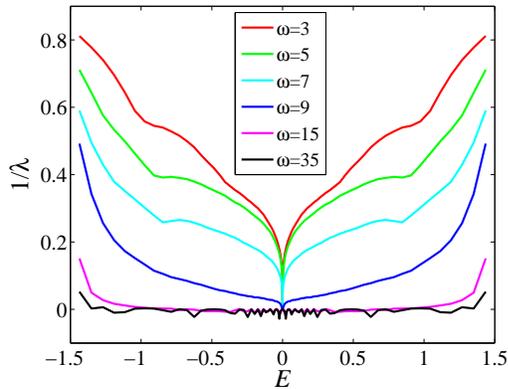}
\caption{Inverse of the localization length $1/\lambda$ vs. quasienergy $E$ (in units of the hopping $J$). Averaged over 1000 realizations of disorder. Hopping amplitude $J=1$, disorde width $\gamma=10J$, oscillation frequency $\omega=3J, 5J, 7J, 9J, 15J, 35J$. }\label{fig:highFreq}
\end{figure}

We have computed the localization length and the density of states for various values of the frequency $\omega$ with fixed hopping amplitude $J=1$ and disorder width $\gamma=10$. The results from the Floquet technique for the original Hamiltonian (\ref{model}) and those obtained from the effective Hamiltonian (\ref{Heff}) at $\omega=3,7,35$ are plotted in FIG. \ref{fig:compare}. One can see that when $\omega=7$ which is comparable to the disorder width $\gamma$ and when $\omega=35$ which is much larger than $\gamma$, the results from the exact Floquet calculation and those from effective Hamiltonian calculation agree quite well. At $\omega=35\gg\gamma=10$, every state is completely delocalized, and the density of states diverges at the two band edges, as expected for a uniform-hopping tight-binding model. On the other hand, at $\omega=7$, both the localization length and the density of states diverge at the band center, which are characteristic of the random hopping model, as expected. The case of $\omega=3$ is slightly more surprising: although the effective Hamiltonian does not work well, the system still exhibits diverging localization length and density of states at the band center. In FIG. \ref{fig:highFreq}, we plot the localization length for more values of $\omega$ from 3 to 35, and the trend from random hopping model behavior to uniform-hopping tight-binding model as $\omega$ increases is clearly seen.

\begin{figure}
\subfigure{\includegraphics[scale = 0.4]{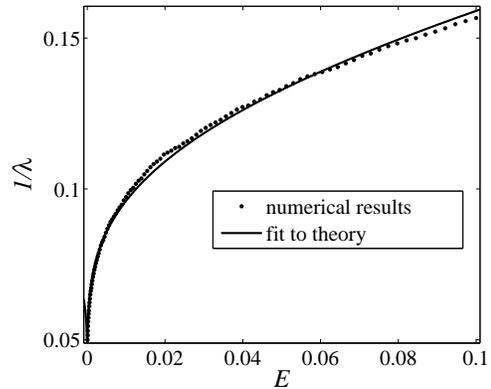}}
\subfigure{\includegraphics[scale = 0.37]{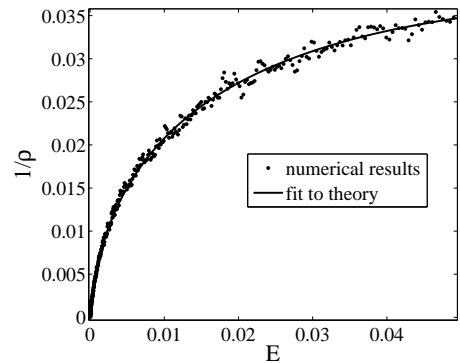}}
\caption{Fitting the inverse of the localization length $\lambda^{-1}$ and the inverse of the density of states $\rho^{-1}$ near the band center to their analytical form 
$
\rho(E) = N\cdot{2\sigma^2}/{|E(\ln (E/E_0)^2)^3|}
$
and
$ 
\lambda(E) = {2|\ln (E/E_0')^2|}/{\sigma^2}
$
. Here, $E_0$ and $E_0'$ are energy scales\cite{Eggarter1978}, and the theoretical value of $\sigma$ is the standard deviation of $ \ln J_{eff}^2$, which is 1.535 for our choice of parameters here. The fitted value of $\sigma$ is 1.496 for $1/\lambda$ and $1.677$ for $1/\rho$. Hopping amplitude $J=1$, oscillation frequency $\omega=7$, and disorder width $\gamma=10$.}\label{fig::closelook}
\end{figure}

Near the band center, we fit the results of localization and the density of states to well-known analytical results\cite{Eggarter1978} (see FIG. \ref{fig::closelook})
\be
\label{Eq:analytical}
\rho(E) &= N\cdot\frac{2\sigma^2}{|E(\ln (E/E_0)^2)^3|},\\
\lambda(E) &= \frac{2|\ln (E/E_0')^2|}{\sigma^2},
\ee
where $N$ is the system size, $E_0$ and $E_0'$ are energy scales, and $\sigma$ is the standard deviation of the logarithm of the effective hopping amplitude square $\ln J_{eff}^2$, with [see Eqn. (\ref{Jeff})]
\be
J_{eff,n}=J\mathcal J_0\left[\frac{2(v_n-v_{n+1})}{\omega}\right].
\ee
We can easily evaluate $\sigma$ numerically to be $1.535$ given $\omega=7$, $\gamma=10$. Fitting numerical results of localization length and density of states, we obtain
\be
\sigma_{fit,\lambda}=1.496, \qquad \sigma_{fit,\rho}=1.677,
\ee
which are quite close to the theoretical value $1.535$ obtained above, further confirming our expectation that random hopping model behavior can be achieved by fast-modulating the onsite energy of Anderson insulators.

\begin{figure}[t]
\centering
\includegraphics[scale = 0.42]{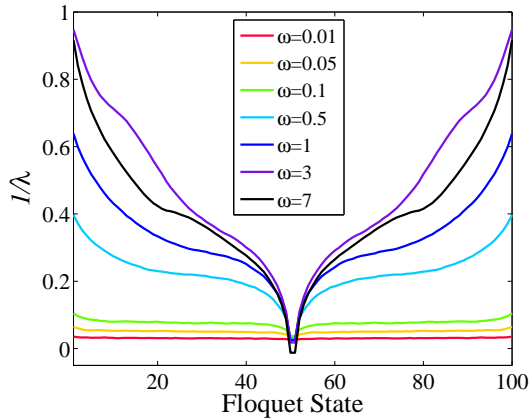}
\caption{(Color online.) Inverse of the localization length $1/\lambda$ vs. the label (e.g., 1st, 2nd, 3rd, ..., 100th) of every Floquet eigenstate for $\omega=0.05,0.1,0.5,1,3,7$. System size $N=100$, hopping amplitude $J=1$, oscillation frequency $\omega=7$, and disorder width $\gamma=10$.}\label{fig:stateNumber}
\end{figure}

At frequencies much smaller than $\gamma$ and the original hopping
strength $J$, interestingly, the system has quite large localization
length in this regime. In FIG. \ref{fig:stateNumber}, we plot the
inverse of the localization length vs. the label (e.g., 1st, 2nd, 3rd,
..., 100th) of every Floquet eigenstate for
$\omega=0.01,0.05,0.1,0.5,1,3,7$ with system size $N=100$ (the total
number of Floquet states equals the system size $N$). One can see that
from $\omega=7$, when $\omega$ is lowered, first the localization
length decreases (inverse of the localization increases), but around
$\omega=3$ this trend is reversed, and all the states become more and
more delocalized at smaller frequencies. At $\omega=0.01,0.05$ and
0.1, all the states have almost equally large localization length.

This trend of delocalization at small frequencies is quite puzzling,
but it is interesting to notice that in a similar model where an
Anderson insulator is manipulated with an AC electric field
\cite{Holthaus1995,Martinez2006r,Martinez2006}, an analogous
delocalization trend was found. An intuitive argument was given in
Ref. \onlinecite{Martinez2006}, where it is argued that the modulation
with frequency $\omega$ allows electrons to absort or emit integer
numbers of ``photons'' with energy $\omega$. When $\omega$ is smaller,
more states with quasienergy $E\pm n\omega$ are in the original energy
band. Since the scale of the localization length of the new state
should be set by the state with the largest localization length among
all the states with energy $E\pm n\omega$, a smaller $\omega$ imples
that it is more likely for the new state to have larger localization
length. This intuitive picture could be of some relevance to our case
here as well.

\section{Experimental Feasibility}\label{sec:experiment}

The main motivation for our work is to experimentally realize the
particle-hole symmetric disorder classes. Following the route outlined
above, requires creating an Anderson insulator in optical lattices,
and then modulate the onsite disorder potential periodically in
time. The behavior of ultracold
atomic systems subject to disorder, and the resulting localization
phenomena is a field of intense current research
\cite{Fallani2008}. Experiments have relied on two methods to
introduce disorder into such systems.  The first involves
using two incommensurate optical lattice potentials, providing
an effective realization of the Aubrey Andre model  which has been
shown to give localization \cite{roati08}.  The second method method uses a 
speckle potential produced by passing a laser through a diffusing
plate which directly imprints a disorder potential \cite{billy08}.

The main challenge in realizing the phenomena introduced in this work
is producing time-dependent disorder potentials which periodically
attract and repell the atoms in the optical lattice system.  The most
direct way to achieve this is to periodically change the detuning of
the disorder potential potential from red to blue.  That is, the 
disorder potential is given by 
\begin{equation}
V(r)=\frac{3\pi c^2}{2\omega_0^2}\left(\frac{\Gamma}{\Delta}\right)I(r),
\end{equation}
where $c$ is the speed of light, $\omega_0$ is the atomic resonance
frequency, $\Delta=\omega-\omega_0$ is the detuning frequency.   
Thus, the sign of the disorder potential can be periodically changed
by periodically  changing the detuning.  This can be achieved by using
an acousto-optic modulator to continuously vary the laser frequency.
However,  sweeping through the responance can produce
undesirable atomic losses.  Thus it might be best to periodically alternate two laser
beams (one red and the other blue detuned) through the same speckle
potential.

The main experimental probe to detect Anderson localization in cold
atom system has been time-of-flight spectroscopy
\cite{roati08,billy08}.  In particular, for weak disorder when the
condensate occupies a delocalized state, the condensate ballistically
expands when the confining potential is removed.  On the other hand
for strong disorder potentials, the condensate occupies localized
states and ceases to expand at a characteristic time after released
from the trap. As we have seen, in the random hopping model 
some of the states are localized while others are delocalized (near
the band center).  Thus disentangling such behavior
using time-of-flight spectroscopy alone might prove to be an
experimental challenge.  On the other hand, the well-developed
technique of Bragg spectroscopy \cite{stenger99,stamper-kurn99} allows the access
to the spectral function and therefore the density of states of
quantum gas.  Thus, perhaps the most promising way of detecting
the Dyson delocalized state is through its distinct single
single-particle density of states near the band center given by
Eq. (\ref{Eq:analytical}) measured through Bragg spectroscopy.

\section{Summary and Discussion}\label{sec:summary}

The simple partical-hole symmetric class of quantum disorder problems exhibits many fascinating
properties such as diverging localization length and diverging density
of states at the band center. It is difficult, however, to realize it experimentally
because its crucial particle-hole symmetry is destroyed by any amount
of diagonal disorder. Our work suggests a realistic path to 
effectively realizing random hopping models in optical lattices by
fast-modulating the disordered potential energies of an Anderson
insulator, but without need for fine tuning the local potential. 
Our idea follows naturally from the recently studied, so-called,
``dynamical localization'' or ``coherent destruction of tunneling''
phenomena as well as from recent efforts to observe Anderson
insulators in optical lattices by using various ways to artificially
produce disordered potentials.

The setup we suggest can explore more than just the static properties
of a special disorder model. After all, we are describing the response
of a quantum system to a strong time dependent potential. By exact
diaonalization of the Floquet operators arising from our model, we
explored the spectral properties of a 1d system subject to strong
time-periodic disorder. As shown in
FIG. \ref{fig:compare} and FIG. \ref{fig:highFreq}, the special
features anticipated from the 1d random hopping problem arise in a
wide range of frequency modulations.  Even with
moderately small frequency $\omega$ where the effective Hamiltonian
(averaged in the vector-potential-only gauge)
does not provide a satisfactory description, the localization length
and the density of states still show the random-hopping-model behavior
(see the case of $\omega=3$ in FIG.\ref{fig:compare}). Furthermore,
a fit of our results near the band-center to the well-known theoretical
form \cite{Eggarter1978} gives good agreement (see
FIG. \ref{fig::closelook}).

Our model, however, gives results which we do not yet intuitively
understand in the low frequency limit. As one can see from
FIG. \ref{fig:highFreq}, the localization length gradually decreases
if the frequency $\omega$ is reduced from 35 to 3 given the disorder
width $\gamma=10$.  When the frequency $\omega$ is further lowered, our
numerical results reveal that the localization length starts to
increase again, and it becomes quite large for $\omega=0.05$ and
$\omega=0.1$ (see FIG. \ref{fig:stateNumber}). We should recall that
the terminus of the limit $\omega\rightarrow 0$ is the well-known
Anderson insulator, in which all states are localized. We intend to
study the localization properties of such Floquet operators with strong
disorder in future work.

While here we concentrated on the realization of particular classes of
quantum disorder problems, the use of time modulated Hamiltonians
could be a general path to the experimental realization of desirable
systems. Indeed, this philosophy is already apparent in the quest to
realize low-filling quantum Hall states in cold atom systems
\cite{DemlerSorensen}, and we expect that in both cold-atom 
and solid-state based devices, time dependent Hamiltonians will
become a standard tool for the stabilization of unique many body wave
functions.

\acknowledgements

For particularly useful discussions we thank  J. Biddle, L. Fallani, S. Rolston,
I. Spielman, and B. Wu. We would like to acknowledge support from the Joint Quantum
Institute Physics Frontier Center and the Sherman Fairchild Foundation
(RB); and the Packard Foundation, Sloan Foundation, and the research
corporation, as well as Darpa, and NSF grants PHY-0456720 and
PHY-0803371 (GR).

\bibliography{reference}

\begin{thebibliography}{44}
\expandafter\ifx\csname natexlab\endcsname\relax\def\natexlab#1{#1}\fi
\expandafter\ifx\csname bibnamefont\endcsname\relax
  \def\bibnamefont#1{#1}\fi
\expandafter\ifx\csname bibfnamefont\endcsname\relax
  \def\bibfnamefont#1{#1}\fi
\expandafter\ifx\csname citenamefont\endcsname\relax
  \def\citenamefont#1{#1}\fi
\expandafter\ifx\csname url\endcsname\relax
  \def\url#1{\texttt{#1}}\fi
\expandafter\ifx\csname urlprefix\endcsname\relax\def\urlprefix{URL }\fi
\providecommand{\bibinfo}[2]{#2}
\providecommand{\eprint}[2][]{\url{#2}}

\bibitem[{\citenamefont{Zirnbauer}(1996)}]{Zirnbauer1996}
\bibinfo{author}{\bibfnamefont{M.~R.} \bibnamefont{Zirnbauer}},
  \bibinfo{journal}{J. Math. Phys.} \textbf{\bibinfo{volume}{37}},
  \bibinfo{pages}{4986} (\bibinfo{year}{1996}).

\bibitem[{\citenamefont{Dyson}(1953)}]{Dyson}
\bibinfo{author}{\bibfnamefont{F.~J.} \bibnamefont{Dyson}},
  \bibinfo{journal}{Phys. Rev.} \textbf{\bibinfo{volume}{92}},
  \bibinfo{pages}{1331} (\bibinfo{year}{1953}).

\bibitem[{\citenamefont{Thouless}(1972)}]{Thouless1972}
\bibinfo{author}{\bibfnamefont{D.~J.} \bibnamefont{Thouless}},
  \bibinfo{journal}{J. Phys. C} \textbf{\bibinfo{volume}{5}},
  \bibinfo{pages}{77} (\bibinfo{year}{1972}).

\bibitem[{\citenamefont{Theodorou and Cohen}(1976)}]{Theodorou1976}
\bibinfo{author}{\bibfnamefont{G.}~\bibnamefont{Theodorou}} \bibnamefont{and}
  \bibinfo{author}{\bibfnamefont{M.~H.} \bibnamefont{Cohen}},
  \bibinfo{journal}{Phys. Rev. B} \textbf{\bibinfo{volume}{13}},
  \bibinfo{pages}{4597} (\bibinfo{year}{1976}).

\bibitem[{\citenamefont{Eggarter and Riedinger}(1978)}]{Eggarter1978}
\bibinfo{author}{\bibfnamefont{T.~P.} \bibnamefont{Eggarter}} \bibnamefont{and}
  \bibinfo{author}{\bibfnamefont{R.}~\bibnamefont{Riedinger}},
  \bibinfo{journal}{Phys. Rev. B} \textbf{\bibinfo{volume}{18}},
  \bibinfo{pages}{569} (\bibinfo{year}{1978}).

\bibitem[{\citenamefont{Fisher}(1994)}]{Fisher1994}
\bibinfo{author}{\bibfnamefont{D.~S.} \bibnamefont{Fisher}},
  \bibinfo{journal}{Phys. Rev. B} \textbf{\bibinfo{volume}{50}},
  \bibinfo{pages}{3799} (\bibinfo{year}{1994}).

\bibitem[{\citenamefont{Balents and Fisher}(1997)}]{Balents1997}
\bibinfo{author}{\bibfnamefont{L.}~\bibnamefont{Balents}} \bibnamefont{and}
  \bibinfo{author}{\bibfnamefont{M.~P.~A.} \bibnamefont{Fisher}},
  \bibinfo{journal}{Phys. Rev. B} \textbf{\bibinfo{volume}{56}},
  \bibinfo{pages}{12970} (\bibinfo{year}{1997}).

\bibitem[{\citenamefont{Gade}(1993)}]{Gade}
\bibinfo{author}{\bibfnamefont{R.}~\bibnamefont{Gade}}, \bibinfo{journal}{Nucl.
  Phys.} \textbf{\bibinfo{volume}{B398}}, \bibinfo{pages}{499}
  (\bibinfo{year}{1993}).

\bibitem[{\citenamefont{Motrunich et~al.}(2002)\citenamefont{Motrunich, Damle,
  and Huse}}]{Motrunich}
\bibinfo{author}{\bibfnamefont{O.}~\bibnamefont{Motrunich}},
  \bibinfo{author}{\bibfnamefont{K.}~\bibnamefont{Damle}}, \bibnamefont{and}
  \bibinfo{author}{\bibfnamefont{D.~A.} \bibnamefont{Huse}},
  \bibinfo{journal}{Phys. Rev. B} \textbf{\bibinfo{volume}{65}},
  \bibinfo{pages}{064206} (\bibinfo{year}{2002}).

\bibitem[{\citenamefont{Foster and Ludwig}(2008)}]{Foster2008}
\bibinfo{author}{\bibfnamefont{M.~S.} \bibnamefont{Foster}} \bibnamefont{and}
  \bibinfo{author}{\bibfnamefont{A.~W.~W.} \bibnamefont{Ludwig}},
  \bibinfo{journal}{Phys. Rev. B} \textbf{\bibinfo{volume}{77}},
  \bibinfo{pages}{165108} (\bibinfo{year}{2008}).

\bibitem[{\citenamefont{Altman et~al.}(2004)\citenamefont{Altman, Kafri,
  Polkovnikov, and Refael}}]{Gil2004}
\bibinfo{author}{\bibfnamefont{E.}~\bibnamefont{Altman}},
  \bibinfo{author}{\bibfnamefont{Y.}~\bibnamefont{Kafri}},
  \bibinfo{author}{\bibfnamefont{A.}~\bibnamefont{Polkovnikov}},
  \bibnamefont{and} \bibinfo{author}{\bibfnamefont{G.}~\bibnamefont{Refael}},
  \bibinfo{journal}{Phys. Rev. Lett.} \textbf{\bibinfo{volume}{93}},
  \bibinfo{pages}{150402} (\bibinfo{year}{2004}).

\bibitem[{\citenamefont{Altman et~al.}(2008)\citenamefont{Altman, Kafri,
  Polkovnikov, and Refael}}]{Gil2008}
\bibinfo{author}{\bibfnamefont{E.}~\bibnamefont{Altman}},
  \bibinfo{author}{\bibfnamefont{Y.}~\bibnamefont{Kafri}},
  \bibinfo{author}{\bibfnamefont{A.}~\bibnamefont{Polkovnikov}},
  \bibnamefont{and} \bibinfo{author}{\bibfnamefont{G.}~\bibnamefont{Refael}},
  \bibinfo{journal}{Phys. Rev. Lett.} \textbf{\bibinfo{volume}{100}},
  \bibinfo{pages}{170402} (\bibinfo{year}{2008}).

\bibitem[{\citenamefont{Dunlap and Kenkre}(1986)}]{Dunlap1986}
\bibinfo{author}{\bibfnamefont{D.~H.} \bibnamefont{Dunlap}} \bibnamefont{and}
  \bibinfo{author}{\bibfnamefont{V.~M.} \bibnamefont{Kenkre}},
  \bibinfo{journal}{Phys. Rev. B} \textbf{\bibinfo{volume}{34}},
  \bibinfo{pages}{3625} (\bibinfo{year}{1986}).

\bibitem[{\citenamefont{Grossmann et~al.}(1991)\citenamefont{Grossmann,
  Dittrich, Jung, and H\"anggi}}]{Grossmann1991}
\bibinfo{author}{\bibfnamefont{F.}~\bibnamefont{Grossmann}},
  \bibinfo{author}{\bibfnamefont{T.}~\bibnamefont{Dittrich}},
  \bibinfo{author}{\bibfnamefont{P.}~\bibnamefont{Jung}}, \bibnamefont{and}
  \bibinfo{author}{\bibfnamefont{P.}~\bibnamefont{H\"anggi}},
  \bibinfo{journal}{Phys. Rev. Lett.} \textbf{\bibinfo{volume}{67}},
  \bibinfo{pages}{516} (\bibinfo{year}{1991}).

\bibitem[{\citenamefont{Holthaus}(1992)}]{Holthaus1992}
\bibinfo{author}{\bibfnamefont{M.}~\bibnamefont{Holthaus}},
  \bibinfo{journal}{Phys. Rev. Lett.} \textbf{\bibinfo{volume}{69}},
  \bibinfo{pages}{351} (\bibinfo{year}{1992}).

\bibitem[{\citenamefont{Ignatov et~al.}(1993)\citenamefont{Ignatov, Renk, and
  Dodin}}]{Ignatov1993}
\bibinfo{author}{\bibfnamefont{A.~A.} \bibnamefont{Ignatov}},
  \bibinfo{author}{\bibfnamefont{K.~F.} \bibnamefont{Renk}}, \bibnamefont{and}
  \bibinfo{author}{\bibfnamefont{E.~P.} \bibnamefont{Dodin}},
  \bibinfo{journal}{Phys. Rev. Lett.} \textbf{\bibinfo{volume}{70}},
  \bibinfo{pages}{1996} (\bibinfo{year}{1993}).

\bibitem[{\citenamefont{Holthaus et~al.}(1995)\citenamefont{Holthaus, Ristow,
  and Hone}}]{Holthaus1995}
\bibinfo{author}{\bibfnamefont{M.}~\bibnamefont{Holthaus}},
  \bibinfo{author}{\bibfnamefont{G.~H.} \bibnamefont{Ristow}},
  \bibnamefont{and} \bibinfo{author}{\bibfnamefont{D.~W.} \bibnamefont{Hone}},
  \bibinfo{journal}{Phys. Rev. Lett.} \textbf{\bibinfo{volume}{75}},
  \bibinfo{pages}{3914} (\bibinfo{year}{1995}).

\bibitem[{\citenamefont{Meier et~al.}(1995)\citenamefont{Meier, von Plessen,
  Thomas, and Koch}}]{Meier1995}
\bibinfo{author}{\bibfnamefont{T.}~\bibnamefont{Meier}},
  \bibinfo{author}{\bibfnamefont{G.}~\bibnamefont{von Plessen}},
  \bibinfo{author}{\bibfnamefont{P.}~\bibnamefont{Thomas}}, \bibnamefont{and}
  \bibinfo{author}{\bibfnamefont{S.~W.} \bibnamefont{Koch}},
  \bibinfo{journal}{Phys. Rev. B} \textbf{\bibinfo{volume}{51}},
  \bibinfo{pages}{14490} (\bibinfo{year}{1995}).

\bibitem[{\citenamefont{Grifoni and Hanggi}(1998)}]{Grifoni}
\bibinfo{author}{\bibfnamefont{M.}~\bibnamefont{Grifoni}} \bibnamefont{and}
  \bibinfo{author}{\bibfnamefont{P.}~\bibnamefont{Hanggi}},
  \bibinfo{journal}{Phys. Rep.} \textbf{\bibinfo{volume}{304}},
  \bibinfo{pages}{229} (\bibinfo{year}{1998}).

\bibitem[{\citenamefont{Holthaus}(2001)}]{Holthaus2001}
\bibinfo{author}{\bibfnamefont{M.}~\bibnamefont{Holthaus}},
  \bibinfo{journal}{Phys. Rev. A} \textbf{\bibinfo{volume}{64}},
  \bibinfo{pages}{011601} (\bibinfo{year}{2001}).

\bibitem[{\citenamefont{Eckardt
  et~al.}(2005{\natexlab{a}})\citenamefont{Eckardt, Weiss, and
  Holthaus}}]{Eckardt2005}
\bibinfo{author}{\bibfnamefont{A.}~\bibnamefont{Eckardt}},
  \bibinfo{author}{\bibfnamefont{C.}~\bibnamefont{Weiss}}, \bibnamefont{and}
  \bibinfo{author}{\bibfnamefont{M.}~\bibnamefont{Holthaus}},
  \bibinfo{journal}{Phys. Rev. Lett.} \textbf{\bibinfo{volume}{95}},
  \bibinfo{pages}{260404} (\bibinfo{year}{2005}{\natexlab{a}}).

\bibitem[{\citenamefont{Eckardt
  et~al.}(2005{\natexlab{b}})\citenamefont{Eckardt, Jinasundera, Weiss, and
  Holthaus}}]{Eckardt2005b}
\bibinfo{author}{\bibfnamefont{A.}~\bibnamefont{Eckardt}},
  \bibinfo{author}{\bibfnamefont{T.}~\bibnamefont{Jinasundera}},
  \bibinfo{author}{\bibfnamefont{C.}~\bibnamefont{Weiss}}, \bibnamefont{and}
  \bibinfo{author}{\bibfnamefont{M.}~\bibnamefont{Holthaus}},
  \bibinfo{journal}{Phys. Rev. Lett.} \textbf{\bibinfo{volume}{95}},
  \bibinfo{pages}{200401} (\bibinfo{year}{2005}{\natexlab{b}}).

\bibitem[{\citenamefont{Creffield and Monteiro}(2006)}]{Creffield2006}
\bibinfo{author}{\bibfnamefont{C.~E.} \bibnamefont{Creffield}}
  \bibnamefont{and} \bibinfo{author}{\bibfnamefont{T.~S.}
  \bibnamefont{Monteiro}}, \bibinfo{journal}{Phys. Rev. Lett.}
  \textbf{\bibinfo{volume}{96}}, \bibinfo{pages}{210403}
  (\bibinfo{year}{2006}).

\bibitem[{\citenamefont{Martinez and
  Molina}(2006{\natexlab{a}})}]{Martinez2006r}
\bibinfo{author}{\bibfnamefont{D.~F.} \bibnamefont{Martinez}} \bibnamefont{and}
  \bibinfo{author}{\bibfnamefont{R.~A.} \bibnamefont{Molina}},
  \bibinfo{journal}{Phys. Rev. B} \textbf{\bibinfo{volume}{73}},
  \bibinfo{pages}{073104} (\bibinfo{year}{2006}{\natexlab{a}}).

\bibitem[{\citenamefont{Martinez and
  Molina}(2006{\natexlab{b}})}]{Martinez2006}
\bibinfo{author}{\bibfnamefont{D.~F.} \bibnamefont{Martinez}} \bibnamefont{and}
  \bibinfo{author}{\bibfnamefont{R.~A.} \bibnamefont{Molina}},
  \bibinfo{journal}{Eur. Phys. J. B.} \textbf{\bibinfo{volume}{52}},
  \bibinfo{pages}{281} (\bibinfo{year}{2006}{\natexlab{b}}).

\bibitem[{\citenamefont{Eckardt and Holthaus}(2007)}]{Eckardt2007}
\bibinfo{author}{\bibfnamefont{A.}~\bibnamefont{Eckardt}} \bibnamefont{and}
  \bibinfo{author}{\bibfnamefont{M.}~\bibnamefont{Holthaus}},
  \bibinfo{journal}{Eur. Phys. Lett.} \textbf{\bibinfo{volume}{80}},
  \bibinfo{pages}{50004} (\bibinfo{year}{2007}).

\bibitem[{\citenamefont{Lignier et~al.}(2007)\citenamefont{Lignier, Sias,
  Ciampini, Singh, Zenesini, Morsch, and Arimondo}}]{Lignier2007}
\bibinfo{author}{\bibfnamefont{H.}~\bibnamefont{Lignier}},
  \bibinfo{author}{\bibfnamefont{C.}~\bibnamefont{Sias}},
  \bibinfo{author}{\bibfnamefont{D.}~\bibnamefont{Ciampini}},
  \bibinfo{author}{\bibfnamefont{Y.}~\bibnamefont{Singh}},
  \bibinfo{author}{\bibfnamefont{A.}~\bibnamefont{Zenesini}},
  \bibinfo{author}{\bibfnamefont{O.}~\bibnamefont{Morsch}}, \bibnamefont{and}
  \bibinfo{author}{\bibfnamefont{E.}~\bibnamefont{Arimondo}},
  \bibinfo{journal}{Phys. Rev. Lett.} \textbf{\bibinfo{volume}{99}},
  \bibinfo{pages}{220403} (\bibinfo{year}{2007}).

\bibitem[{\citenamefont{Kierig et~al.}(2008)\citenamefont{Kierig,
  Schnorrberger, Schietinger, Tomkovic, and Oberthaler}}]{Kierig2008}
\bibinfo{author}{\bibfnamefont{E.}~\bibnamefont{Kierig}},
  \bibinfo{author}{\bibfnamefont{U.}~\bibnamefont{Schnorrberger}},
  \bibinfo{author}{\bibfnamefont{A.}~\bibnamefont{Schietinger}},
  \bibinfo{author}{\bibfnamefont{J.}~\bibnamefont{Tomkovic}}, \bibnamefont{and}
  \bibinfo{author}{\bibfnamefont{M.~K.} \bibnamefont{Oberthaler}},
  \bibinfo{journal}{Phys. Rev. Lett.} \textbf{\bibinfo{volume}{100}},
  \bibinfo{pages}{190405} (\bibinfo{year}{2008}).

\bibitem[{\citenamefont{Luo et~al.}(2008)\citenamefont{Luo, Xie, and
  Wu}}]{Luo2008}
\bibinfo{author}{\bibfnamefont{X.}~\bibnamefont{Luo}},
  \bibinfo{author}{\bibfnamefont{Q.}~\bibnamefont{Xie}}, \bibnamefont{and}
  \bibinfo{author}{\bibfnamefont{B.}~\bibnamefont{Wu}}, \bibinfo{journal}{Phys.
  Rev. A} \textbf{\bibinfo{volume}{77}}, \bibinfo{pages}{053601}
  (\bibinfo{year}{2008}).

\bibitem[{\citenamefont{Tsukada et~al.}(2008)\citenamefont{Tsukada, Yoshida,
  and Suzuki}}]{Tsukada2008}
\bibinfo{author}{\bibfnamefont{N.}~\bibnamefont{Tsukada}},
  \bibinfo{author}{\bibfnamefont{H.}~\bibnamefont{Yoshida}}, \bibnamefont{and}
  \bibinfo{author}{\bibfnamefont{T.}~\bibnamefont{Suzuki}},
  \bibinfo{journal}{Phys. Rev. A} \textbf{\bibinfo{volume}{77}},
  \bibinfo{pages}{022101} (\bibinfo{year}{2008}).

\bibitem[{\citenamefont{Kayanuma and Saito}(2008)}]{Kayanuma}
\bibinfo{author}{\bibfnamefont{Y.}~\bibnamefont{Kayanuma}} \bibnamefont{and}
  \bibinfo{author}{\bibfnamefont{K.}~\bibnamefont{Saito}},
  \bibinfo{journal}{Phys. Rev. A} \textbf{\bibinfo{volume}{77}},
  \bibinfo{pages}{010101} (\bibinfo{year}{2008}).

\bibitem[{\citenamefont{Creffield}(2009)}]{Creffield2009}
\bibinfo{author}{\bibfnamefont{C.~E.} \bibnamefont{Creffield}},
  \bibinfo{journal}{Phys. Rev. A} \textbf{\bibinfo{volume}{79}},
  \bibinfo{pages}{063612} (\bibinfo{year}{2009}).

\bibitem[{\citenamefont{Eckardt et~al.}(2009)\citenamefont{Eckardt, Holthaus,
  Lignier, Zenesini, Ciampini, Morsch, and Arimondo}}]{Eckardt2009}
\bibinfo{author}{\bibfnamefont{A.}~\bibnamefont{Eckardt}},
  \bibinfo{author}{\bibfnamefont{M.}~\bibnamefont{Holthaus}},
  \bibinfo{author}{\bibfnamefont{H.}~\bibnamefont{Lignier}},
  \bibinfo{author}{\bibfnamefont{A.}~\bibnamefont{Zenesini}},
  \bibinfo{author}{\bibfnamefont{D.}~\bibnamefont{Ciampini}},
  \bibinfo{author}{\bibfnamefont{O.}~\bibnamefont{Morsch}}, \bibnamefont{and}
  \bibinfo{author}{\bibfnamefont{E.}~\bibnamefont{Arimondo}},
  \bibinfo{journal}{Phys. Rev. A} \textbf{\bibinfo{volume}{79}},
  \bibinfo{pages}{013611} (\bibinfo{year}{2009}).

\bibitem[{\citenamefont{Sias et~al.}(2008)\citenamefont{Sias, Lignier, Singh,
  Zenesini, Ciampini, Morsch, and Arimondo}}]{Sias}
\bibinfo{author}{\bibfnamefont{C.}~\bibnamefont{Sias}},
  \bibinfo{author}{\bibfnamefont{H.}~\bibnamefont{Lignier}},
  \bibinfo{author}{\bibfnamefont{Y.~P.} \bibnamefont{Singh}},
  \bibinfo{author}{\bibfnamefont{A.}~\bibnamefont{Zenesini}},
  \bibinfo{author}{\bibfnamefont{D.}~\bibnamefont{Ciampini}},
  \bibinfo{author}{\bibfnamefont{O.}~\bibnamefont{Morsch}}, \bibnamefont{and}
  \bibinfo{author}{\bibfnamefont{E.}~\bibnamefont{Arimondo}},
  \bibinfo{journal}{Phys. Rev. Lett.} \textbf{\bibinfo{volume}{100}},
  \bibinfo{pages}{040404} (\bibinfo{year}{2008}).

\bibitem[{\citenamefont{Zenesini et~al.}(2009)\citenamefont{Zenesini, Lignier,
  Ciampini, Morsch, and Arimondo}}]{Zenesini}
\bibinfo{author}{\bibfnamefont{A.}~\bibnamefont{Zenesini}},
  \bibinfo{author}{\bibfnamefont{H.}~\bibnamefont{Lignier}},
  \bibinfo{author}{\bibfnamefont{D.}~\bibnamefont{Ciampini}},
  \bibinfo{author}{\bibfnamefont{O.}~\bibnamefont{Morsch}}, \bibnamefont{and}
  \bibinfo{author}{\bibfnamefont{E.}~\bibnamefont{Arimondo}},
  \bibinfo{journal}{Phys. Rev. Lett.} \textbf{\bibinfo{volume}{102}},
  \bibinfo{pages}{100403} (\bibinfo{year}{2009}).

\bibitem[{\citenamefont{Shirley}(1965)}]{Shirley}
\bibinfo{author}{\bibfnamefont{J.~H.} \bibnamefont{Shirley}},
  \bibinfo{journal}{Phys. Rev.} \textbf{\bibinfo{volume}{138}},
  \bibinfo{pages}{B979} (\bibinfo{year}{1965}).

\bibitem[{\citenamefont{Sambe}(1973)}]{Sambe1973}
\bibinfo{author}{\bibfnamefont{H.}~\bibnamefont{Sambe}},
  \bibinfo{journal}{Phys. Rev. A} \textbf{\bibinfo{volume}{7}},
  \bibinfo{pages}{2203} (\bibinfo{year}{1973}).

\bibitem[{\citenamefont{Martinez}(2003)}]{Martinez2003}
\bibinfo{author}{\bibfnamefont{D.~F.} \bibnamefont{Martinez}},
  \bibinfo{journal}{J. Phys. A: Math. Gen.} \textbf{\bibinfo{volume}{36}},
  \bibinfo{pages}{9827} (\bibinfo{year}{2003}).

\bibitem[{\citenamefont{Fallani et~al.}()\citenamefont{Fallani, Fort, and
  Inguscio}}]{Fallani2008}
\bibinfo{author}{\bibfnamefont{L.}~\bibnamefont{Fallani}},
  \bibinfo{author}{\bibfnamefont{C.}~\bibnamefont{Fort}}, \bibnamefont{and}
  \bibinfo{author}{\bibfnamefont{M.}~\bibnamefont{Inguscio}},
  \bibinfo{note}{arXiv:0804.2888}.

\bibitem[{\citenamefont{Roati et~al.}(2008)\citenamefont{Roati, D'Errico,
  Fallani, Fattori, Fort, Zaccanti, Modugno, Modugno, and Inguscio}}]{roati08}
\bibinfo{author}{\bibfnamefont{G.}~\bibnamefont{Roati}},
  \bibinfo{author}{\bibfnamefont{C.}~\bibnamefont{D'Errico}},
  \bibinfo{author}{\bibfnamefont{L.}~\bibnamefont{Fallani}},
  \bibinfo{author}{\bibfnamefont{M.}~\bibnamefont{Fattori}},
  \bibinfo{author}{\bibfnamefont{C.}~\bibnamefont{Fort}},
  \bibinfo{author}{\bibfnamefont{M.}~\bibnamefont{Zaccanti}},
  \bibinfo{author}{\bibfnamefont{G.}~\bibnamefont{Modugno}},
  \bibinfo{author}{\bibfnamefont{M.}~\bibnamefont{Modugno}}, \bibnamefont{and}
  \bibinfo{author}{\bibfnamefont{M.}~\bibnamefont{Inguscio}},
  \bibinfo{journal}{Nature} \textbf{\bibinfo{volume}{453}},
  \bibinfo{pages}{895} (\bibinfo{year}{2008}).

\bibitem[{\citenamefont{Billy et~al.}(2008)\citenamefont{Billy, Josse, Zuo,
  Bernard, Hambrecht, Lugan, Clément, Sanchez-Palencia, Bouyer, and
  Aspect}}]{billy08}
\bibinfo{author}{\bibfnamefont{J.}~\bibnamefont{Billy}},
  \bibinfo{author}{\bibfnamefont{V.}~\bibnamefont{Josse}},
  \bibinfo{author}{\bibfnamefont{Z.}~\bibnamefont{Zuo}},
  \bibinfo{author}{\bibfnamefont{A.}~\bibnamefont{Bernard}},
  \bibinfo{author}{\bibfnamefont{B.}~\bibnamefont{Hambrecht}},
  \bibinfo{author}{\bibfnamefont{P.}~\bibnamefont{Lugan}},
  \bibinfo{author}{\bibfnamefont{D.}~\bibnamefont{Clément}},
  \bibinfo{author}{\bibfnamefont{L.}~\bibnamefont{Sanchez-Palencia}},
  \bibinfo{author}{\bibfnamefont{P.}~\bibnamefont{Bouyer}}, \bibnamefont{and}
  \bibinfo{author}{\bibfnamefont{A.}~\bibnamefont{Aspect}},
  \bibinfo{journal}{Nature} \textbf{\bibinfo{volume}{453}},
  \bibinfo{pages}{891} (\bibinfo{year}{2008}).

\bibitem[{\citenamefont{Stenger et~al.}(1999)\citenamefont{Stenger, Inouye,
  Chikkatur, Stamper-Kurn, Pritchard, and Ketterle}}]{stenger99}
\bibinfo{author}{\bibfnamefont{J.}~\bibnamefont{Stenger}},
  \bibinfo{author}{\bibfnamefont{S.}~\bibnamefont{Inouye}},
  \bibinfo{author}{\bibfnamefont{A.~P.} \bibnamefont{Chikkatur}},
  \bibinfo{author}{\bibfnamefont{D.~M.} \bibnamefont{Stamper-Kurn}},
  \bibinfo{author}{\bibfnamefont{D.~E.} \bibnamefont{Pritchard}},
  \bibnamefont{and} \bibinfo{author}{\bibfnamefont{W.}~\bibnamefont{Ketterle}},
  \bibinfo{journal}{Phys. Rev. Lett.} \textbf{\bibinfo{volume}{82}},
  \bibinfo{pages}{4569} (\bibinfo{year}{1999}).

\bibitem[{\citenamefont{Stamper-Kurn et~al.}(1999)\citenamefont{Stamper-Kurn,
  Chikkatur, G\"orlitz, Inouye, Gupta, Pritchard, and
  Ketterle}}]{stamper-kurn99}
\bibinfo{author}{\bibfnamefont{D.~M.} \bibnamefont{Stamper-Kurn}},
  \bibinfo{author}{\bibfnamefont{A.~P.} \bibnamefont{Chikkatur}},
  \bibinfo{author}{\bibfnamefont{A.}~\bibnamefont{G\"orlitz}},
  \bibinfo{author}{\bibfnamefont{S.}~\bibnamefont{Inouye}},
  \bibinfo{author}{\bibfnamefont{S.}~\bibnamefont{Gupta}},
  \bibinfo{author}{\bibfnamefont{D.~E.} \bibnamefont{Pritchard}},
  \bibnamefont{and} \bibinfo{author}{\bibfnamefont{W.}~\bibnamefont{Ketterle}},
  \bibinfo{journal}{Phys. Rev. Lett.} \textbf{\bibinfo{volume}{83}},
  \bibinfo{pages}{2876} (\bibinfo{year}{1999}).

\bibitem[{\citenamefont{S\o{}rensen et~al.}(2005)\citenamefont{S\o{}rensen,
  Demler, and Lukin}}]{DemlerSorensen}
\bibinfo{author}{\bibfnamefont{A.~S.} \bibnamefont{S\o{}rensen}},
  \bibinfo{author}{\bibfnamefont{E.}~\bibnamefont{Demler}}, \bibnamefont{and}
  \bibinfo{author}{\bibfnamefont{M.~D.} \bibnamefont{Lukin}},
  \bibinfo{journal}{Phys. Rev. Lett.} \textbf{\bibinfo{volume}{94}},
  \bibinfo{pages}{086803} (\bibinfo{year}{2005}).

\end{thebibliography}
\end{document}